\documentclass[english]{article}%
\usepackage[T1]{fontenc}
\usepackage[latin1]{inputenc}
\usepackage{amsmath}
\usepackage{amssymb}
\usepackage{amsfonts}
\usepackage{babel}
\usepackage{graphicx}%
\makeatletter
\newtheorem{theorem}{Theorem}
\newtheorem{acknowledgement}[theorem]{Acknowledgement}

\makeatother
\begin{document}

\title{{\LARGE Two and four-level systems in magnetic fields restricted in time}}
\author{M.C. Baldiotti\thanks{E-mail: baldiott@fma.if.usp.br}, V.G. Bagrov\thanks{On
leave of absence from Tomsk State University and Tomsk Institute of High
Current Electronics, Russia, e-mail: bagrov@phys.tsu.ru}, D.M.
Gitman\thanks{E-mail: gitman@dfn.if.usp.br}, and A.D. Levin\thanks{Dexter
Research Center, USA; e-mail: SLevin@dexterresearch.com}\\Instituto de Física, Universidade de São Paulo,\\Caixa Postal 66318-CEP, 05315-970 São Paulo, S.P., Brazil}
\maketitle

\begin{abstract}
We describe some new exact solutions for two- and four-level systems. In all
the cases, external fields have a restricted behavior in time. First, we
consider a method to construct new solutions for one-spin equation and give
some explicit examples, one of them is in a external magnetic field that acts
during a finite time interval. Then we show how these solutions can be used to
solve the two-spin equation problem. A solution for two interacting spins is
analized in the case when the field difference between the external fields in
each spin vary adiabatically, vanishing on the time infinity. The latter
system can be identified with a quantum gate realized by two coupled quantum
dots. The probability of the Swap operation for such a gate can be explicitly
expressed in terms of special functions. Using the obtained expressions, we
construct plots for the Swap operation for some parameters of the external
magnetic field and interaction function.

\end{abstract}

\section{Introduction}

Finite-level systems have always played an important role in quantum physics.
In particular, two-level systems possess a wide range of applications, for
example, in the semi-classical theory of laser beams \cite{Nus73}, optical
resonance \cite{AllEb75}, and nuclear induction experiments \cite{RabRaS54},
and so on. The best known physical system that could be identified with a
two-level system is a fixed spin-one-half object interacting with a magnetic
field. The two-level system have been studied by many authors using different
methods, see for example \cite{BarCo02, ShaSaG07}. Likewise the four-level
systems can be used to describe two interacting one-half spins, e.g., the
valence electrons in two coupled semiconductor quantum dots \cite{LosDi98}.
The most detailed theoretical study of the quantum mechanical equations for
two and four-level systems, and their exact solutions, are presented in
\cite{BagGiBL05, BagBaGL07}. Recently, two- and four-level systems have
attract even more attention, due to their relationship to the problem of
quantum computation \cite{NieCh00} . In this problem, the computation is
performed by the manipulation of the so-called one- and two-qubit gates
\cite{BreDaDGHMNO02}. The one-qubit gate can be identified with a two-level
system and two-qubit gates can be identified with a four-level system. For
these reasons, two- and four-level systems are crucial elements of possible
quantum computers, which are supposed to efficiently solve problems that are
considered intractable by classical computers \cite{Shor94,Fey82}. For
physical applications, it is very important to have explicit exact solutions
of two- and four-level system equations. In \cite{BalBaG09,BalGi07} exact
solutions of a four-level system are used to describe the theoretical
construction of a universal quantum XOR gate using two-coupled quantum dots.
This work shows how the exact solutions can be used to establish all the
necessary conditions on the external fields needed for the implementation of
the gate.

In the present work, we describe some new exact solutions for two- and
four-level systems that were not represented in our previous works
\cite{BagGiBL05,BagBaGL07}. These solutions are found for external fields that
have a restricted behavior in time, for example, the first solution for
two-level system in external field that acts along a finite time interval. In
Section 2 we describe a general method to construct exact\textbf{ }solutions
for two-level systems with external fields restricted in time\textbf{, }and
also in this section, we use this method to obtain two explicit external
fields and the respective exact solutions. In Section 3 we show how these
results can be applied to construct new exact solutions for the four-level
system. This system is identified with two interacting spins and we made a
detailed study of the important case when the difference between the external
fields in each spin vary adiabatically (vanishes with time). This system can
be identified with a quantum gate realized by two coupled quantum dots. The
probability of the Swap operation for such a gate can be explicitly expressed
in terms of special functions. Using the obtained expressions, we construct
plots for the Swap operation for some parameters of the external magnetic
field and interaction functions.

\section{Two-level systems}

\subsection{General}

We recall to the reader that two-level systems are described by the so-called
one-spin equation
\begin{equation}
i\frac{dv}{dt}=\left(  \boldsymbol{\sigma}\mathbf{F}\right)
v,\ \boldsymbol{\sigma}=\left(  \sigma_{1},\sigma_{2},\sigma_{3}\right)
,\ \mathbf{F}=\left(  F_{1},F_{2},F_{3}\right)  , \label{1a}%
\end{equation}
where $v=v\left(  t\right)  $ is a two--component spinor, $\sigma_{k}~\left(
k=1,2,3\right)  $\ are Pauli matrices, $F_{k}=F_{k}\left(  t\right)  $ are
components of external field strength, see \cite{BagGiBL05}. The general
solution of the spin equation reads
\begin{equation}
v\left(  t\right)  =R\left(  t\right)  v^{0}, \label{2a}%
\end{equation}
where the $2\times2$ matrix $R\left(  t\right)  $ obeys the same spin
equation
\begin{equation}
i\frac{dR\left(  t\right)  }{dt}=\left(  \boldsymbol{\sigma}\mathbf{F}\right)
R\left(  t\right)  , \label{3a}%
\end{equation}
and $v^{0}$ is an arbitrary constant spinor. If $R\left(  t=t_{0}\right)
=I$,\thinspace where $I$ is $2\times2$ unity matrix, then $R\left(  t\right)
=\hat{u}\left(  t\right)  $, where $\hat{u}\left(  t\right)  $ is the
evolution operator of the spin equation. In the general case, the evolution
operator is constructed by the help of the matrix $R\left(  t\right)  $ as
follows
\begin{equation}
\hat{u}\left(  t\right)  =R\left(  t\right)  R_{0}^{-1},\ R_{0}=R\left(
t=t_{0}\right)  ,\ \hat{u}\left(  t=t_{0}\right)  =I~. \label{4a}%
\end{equation}

The matrix $R(t)$ can always be represented in the form
\begin{equation}
R\left(  t\right)  =Ip_{0}-i\left(  \boldsymbol{\sigma}\mathbf{p}\right)
,\ \mathbf{p}=\left(  p_{1},p_{2},p_{3}\right)  ,\ p_{s}=p_{s}\left(
t\right)  ,\ s=0,1,2,3~. \label{6a}%
\end{equation}
The functions $p_{s}(t)$ obey the set of equations
\begin{equation}
\dot{p}_{0}+\left(  \mathbf{pF}\right)  =0,\ \mathbf{\dot{p}}+\left[
\mathbf{p\times F}\right]  -p_{0}\mathbf{F}=0~, \label{7a}%
\end{equation}
which follows from (\ref{3a}). Equations (\ref{7a}) imply that $\Delta=\det
R\left(  t\right)  =p_{0}^{2}+\mathbf{p}^{2}$ is an integral of motion.

Let us suppose that spin equation (\ref{3a}) is self-adjoint, which means that
the external field $\mathbf{F}$ is real. In such a case we can chose the
functions $p_{s}\left(  t\right)  $ to be real. Without loss of generality, in
such a case, we can set $\Delta=1$, which means that $R\left(  t\right)  $ is
nonsingular. Under the condition $\Delta=1$, the functions $p_{s}\left(
t\right)  $ can be expressed via three real parameters $\alpha=\alpha\left(
t\right)  ,\,\theta=\theta\left(  t\right)  ,$ and $\,\varphi=\varphi\left(
t\right)  $ as follows%

\begin{align}
p_{0}  &  =\cos\frac{\varphi-\alpha}{2}\cos\frac{\theta}{2}~,\ p_{1}%
=-\sin\frac{\varphi+\alpha}{2}\sin\frac{\theta}{2}~,\nonumber\\
p_{2}  &  =\cos\frac{\varphi+\alpha}{2}\sin\frac{\theta}{2}~,\ p_{3}=\sin
\frac{\varphi-\alpha}{2}\cos\frac{\theta}{2}~. \label{8a}%
\end{align}
With these functions, the evolution operator (\ref{6a}) assumes the
form\footnote{The operators $R_{i}\left(  \beta\right)  $ in (\ref{8c}) are
rotation of an angle $\beta$ along the $i$-axis and the functions
$\varphi,\theta\,,\alpha$ can be identify with the Euler angles. In a similar
manner, the functions $p_{i}$\ in (\ref{8a}) can be identify with the Euler
parameters.}%
\begin{equation}
R=R_{3}\left(  -\varphi\right)  R_{2}\left(  \theta\right)  R_{3}\left(
\alpha\right)  ~,\ R_{i}\left[  \beta\left(  t\right)  \right]  =\exp\left[
i\sigma_{i}\frac{\beta\left(  t\right)  }{2}\right]  ~. \label{8c}%
\end{equation}
Substituting this expression in (\ref{3a}) we see that the above $R\left(
t\right)  $ is the evolution operator for the spin equation in the following
external field%
\begin{align}
F_{1}  &  =\frac{\dot{\theta}}{2}\sin\varphi+\frac{\dot{\alpha}}{2}\sin
\theta\cos\varphi~,\nonumber\\
F_{2}  &  =\frac{\dot{\alpha}}{2}\sin\theta\sin\varphi-\frac{\dot{\theta}}%
{2}\cos\varphi~,\nonumber\\
F_{3}  &  =\frac{\dot{\varphi}}{2}-\frac{\dot{\alpha}}{2}\cos\theta~.
\label{8b}%
\end{align}
For any continuous time-dependent functions $\varphi,\theta,\alpha$ with
continuous derivative.

Of course, the above expressions can not be used to solve the general problem
(\ref{1a})\ for a an arbitrary general field $\mathbf{F}$, due to the
difficulty to solve the integral equations involved in to express $\left(
\varphi,\theta,\alpha\right)  $ as a function of $\left(  F_{1},F_{2}%
,F_{3}\right)  $. However, these expressions can be used to find exact
solutions for external fields with some particular characteristic. For
example, we can construct exactly solutions for periodic external fields by
setting%
\[
\theta=\omega t~,\ \alpha=\Omega t~,\ \cos\varphi=C~,
\]
where $\omega,\Omega$ and $C$ are constants. Other important kind of external
fields are those whose action is restricted in time, once these are the most
common fields in experiments. We can find a variety of solutions for restrict
in time external fields, just by constructing functions that assume a constant
value outside of some interval. In the next section we give some explicit
realization of this kind of fields.

\subsection{Exact solutions for some restricted in time external fields}

We can use the results of the above section to construct exact solution of the
spin equation\ for external fields restricted in time. It can be done by
constructing continuous time-dependent functions $\varphi,\theta,\alpha$, with
continuous derivatives, which assumes a fixed value outside of a certain
interval ($\dot{\varphi}=\dot{\theta}=\dot{\alpha}=0\ $for $\left\vert
t\right\vert >T$).

\begin{enumerate}
\item Let us chose the functions:%
\begin{align}
\theta\left(  t\right)   &  =\theta_{0}~,\ \varphi\left(  t\right)
=\varphi_{0},~\alpha\left(  t\right)  =\alpha_{0}~,\ t\leq-T~,\nonumber\\
\theta\left(  t\right)   &  =\frac{\theta_{1}-\theta_{0}}{2}\sin\frac{\pi
t}{2T}+\frac{\theta_{1}+\theta_{0}}{2}~,\ \left\vert t\right\vert
<T~,\nonumber\\
\varphi\left(  t\right)   &  =\frac{\varphi_{1}-\varphi_{0}}{2}\sin\frac{\pi
t}{2T}+\frac{\varphi_{1}+\varphi_{0}}{2}~,\ \left\vert t\right\vert
<T~,\nonumber\\
\alpha\left(  t\right)   &  =\frac{\alpha_{1}-\alpha_{0}}{2}\sin\frac{\pi
t}{2T}+\frac{\alpha_{1}+\alpha_{0}}{2}~,\ \left\vert t\right\vert
<T~,\nonumber\\
\theta\left(  t\right)   &  =\theta_{1}~,\ \varphi\left(  t\right)
=\varphi_{1}~,\alpha\left(  t\right)  =\alpha_{1}~,\ t\geq T~, \label{13}%
\end{align}
where $\alpha_{0},\alpha_{1},\theta_{0},\theta_{1},\varphi_{0},$ and
$\varphi_{1}$ are arbitrary constants. With this, the external field
$\mathbf{F}$\ (\ref{8b})\ will be zero at $|t|\geq T$, where $T$\textrm{\ }is
a constant, and for $|t|<T$ we have
\begin{align*}
F_{1}\left(  t\right)   &  =\frac{\pi}{8T}\left(  \theta_{0}-\theta
_{1}\right)  \cos\frac{\pi t}{2T}\sin\varphi+\frac{\pi}{8T}\left(  \alpha
_{0}-\alpha_{1}\right)  \cos\frac{\pi t}{2T}\sin\theta\cos\varphi,\\
F_{2}\left(  t\right)   &  =\frac{\pi}{8T}\left(  \alpha_{0}-\alpha
_{1}\right)  \cos\frac{\pi t}{2T}\sin\theta\sin\varphi-\frac{\pi}{8T}\left(
\theta_{0}-\theta_{1}\right)  \cos\frac{\pi t}{2T}\cos\varphi,\\
F_{3}\left(  t\right)   &  =\frac{\pi}{8T}\left(  \varphi_{1}-\varphi
_{0}\right)  \cos\frac{\pi t}{2T}-\frac{\pi}{8T}\left(  \alpha_{0}-\alpha
_{1}\right)  \cos\frac{\pi t}{2T}\cos\theta.
\end{align*}
The external field under consideration is not zero only on a finite interval
$\left\vert t\right\vert <T$ and is continuous for all $t$. The exact solution
of the equation (\ref{3a}) for such a field is constructed by substituting the
function (\ref{13}) in (\ref{8a}).

\item Let the functions $\theta=\theta(t)$, $\varphi=\varphi(t)$ and
$\alpha\left(  t\right)  $ have the form%
\begin{align}
\theta(t)  &  =\frac{\theta_{0}t}{T_{1}}\exp\left[  -\left(  \frac{t}{T_{1}%
}\right)  ^{2}\right]  +\theta_{1}~,\ \varphi(t)=\frac{\varphi_{0}t}{T_{2}%
}\exp\left[  -\left(  \frac{t}{T_{2}}\right)  ^{2}\right]  +\varphi
_{1}~,\nonumber\\
\alpha\left(  t\right)   &  =\frac{\alpha_{0}t}{T_{3}}\exp\left[  -\left(
\frac{t}{T_{3}}\right)  ^{2}\right]  +\alpha_{1}~, \label{14}%
\end{align}
where $\alpha_{1},\theta_{0},\,\theta_{1},\,\varphi_{0},\,\varphi_{1}%
,\,\alpha_{0},\,\ $and $T_{k},~\left(  k=1,2,3\right)  $ are arbitrary
constants. With this, the external field $\mathbf{F}$\ (\ref{8b}) becomes
\begin{align}
F_{1}\left(  t\right)   &  =-\frac{\theta_{0}}{T_{1}}\left[  1-2\left(
\frac{t}{T_{1}}\right)  ^{2}\right]  \exp\left[  -\left(  \frac{t}{T_{1}%
}\right)  ^{2}\right]  \sin\varphi\nonumber\\
&  -\frac{\alpha_{0}}{T_{3}}\left[  1-2\left(  \frac{t}{T_{3}}\right)
^{2}\right]  \exp\left[  -\left(  \frac{t}{T_{3}}\right)  ^{2}\right]
\sin\theta\cos\varphi~,\nonumber\\
F_{2}\left(  t\right)   &  =\frac{\theta_{0}}{T_{1}}\left[  1-2\left(
\frac{t}{T_{1}}\right)  ^{2}\right]  \exp\left[  -\left(  \frac{t}{T_{1}%
}\right)  ^{2}\right]  \cos\varphi\nonumber\\
&  -\frac{\alpha_{0}}{T_{3}}\left[  1-2\left(  \frac{t}{T_{3}}\right)
^{2}\right]  \exp\left[  -\left(  \frac{t}{T_{3}}\right)  ^{2}\right]
\sin\theta\sin\varphi~,\nonumber\\
F_{3}\left(  t\right)   &  =\frac{\varphi_{0}}{T_{2}}\left[  1-2\left(
\frac{t}{T_{2}}\right)  ^{2}\right]  \exp\left[  -\left(  \frac{t}{T_{2}%
}\right)  ^{2}\right] \nonumber\\
&  +\frac{\alpha_{0}}{T_{3}}\left[  1-2\left(  \frac{t}{T_{3}}\right)
^{2}\right]  \exp\left[  -\left(  \frac{t}{T_{3}}\right)  ^{2}\right]
\cos\theta~. \label{16a}%
\end{align}
This external field vanishes at $t\rightarrow\pm\infty$. The exact solution of
the equation (\ref{3a}) for such a field is constructed by substituting the
function (\ref{14}) in (\ref{8a}).

\item By combining the $\varphi,\theta,\alpha$ functions of the two above
examples it is possible to construct $6$ fields more in the form (\ref{8b})
that are restricted in time, with the exact solutions given by (\ref{6a}).
\end{enumerate}

\section{Four-level systems}

\subsection{General}

We write the Schrödinger equation for a four-level system in the following
form ($\hbar=1$), see \cite{BagBaGL07}:%
\begin{align}
&  i\frac{d\Psi}{dt}=\hat{H}\left(  \mathbf{G,F,}J\right)  \Psi\,,\nonumber\\
&  \hat{H}=\left(  \boldsymbol{\rho}\mathbf{\cdot G}\right)  +\left(
\boldsymbol{\Sigma}\mathbf{\cdot F}\right)  +\frac{J}{2}\left(
\boldsymbol{\Sigma}\mathbf{\cdot}\boldsymbol{\rho}\right)  . \label{1}%
\end{align}
Here $\Psi$ is a four-component column; in the general case the interaction
function $J$, as well as, the external fields (the vectors $\mathbf{G}$ and
$\mathbf{F}$) are time-dependent; and the $4\times4$\ matrices
$\boldsymbol{\rho}$ and $\boldsymbol{\Sigma}$ have the forms
\[
\boldsymbol{\Sigma}=I\otimes\boldsymbol{\sigma}\,,\;\boldsymbol{\rho
}=\boldsymbol{\sigma}\otimes I\,,\left(  \boldsymbol{\Sigma}\cdot
\boldsymbol{\rho}\right)  =\boldsymbol{\sigma}\otimes\boldsymbol{\sigma}%
=\sum_{i=1}^{3}\sigma_{i}\otimes\sigma_{i}\,,
\]
where $\boldsymbol{\sigma}=\left(  \sigma_{1},\sigma_{2},\sigma_{3}\right)  $
are the Pauli matrices, and $I$ is the $2\times2$ identity matrices. The
Hamiltonian matrix reads%
\begin{equation}
\hat{H}=\left(
\begin{array}
[c]{cccc}%
F_{3}+G_{3}+\frac{J}{2} & F_{1}-iF_{2} & G_{1}-iG_{2} & 0\\
F_{1}+iF_{2} & G_{3}-F_{3}-\frac{J}{2} & J & G_{1}-iG_{2}\\
G_{1}+iG_{2} & J & F_{3}-G_{3}-\frac{J}{2} & F_{1}-iF_{2}\\
0 & G_{1}+iG_{2} & F_{1}+iF_{2} & \frac{J}{2}-G_{3}-F_{3}%
\end{array}
\right)  ~. \label{1b}%
\end{equation}

Such a model is used to describe two spins subjected to the external magnetic
fields $\mathbf{F}$ and $\mathbf{G}$, and interacting with each other through
a spherically symmetric Heisenberg interaction whose intensity is given by the
interaction function $J$. In particular, this model was used to describe two
coupled quantum dots \cite{BalGi07}. In our work \cite{BagBaGL07} a series of
exact solution of equation (\ref{1}) for different choices of the interaction
function and the external fields are found for the first time.

\subsection{Reduction to the two-level system case}

For a special case of two spins subjected to \textit{parallel} external
magnetic fields, which we write as%
\begin{equation}
\mathbf{G}=\left(  0,0,\mu_{B}g_{1}B_{1}\right)  \,,\;\mathbf{F}=\left(
0,0,\mu_{B}g_{2}B_{2}\right)  ~,\;B_{1,2}=B_{1,2}\left(  t\right)
\mathrm{\,}, \label{5}%
\end{equation}
where $\mu_{B}$ is the Bohr magneton and $g_{1}$ and $g_{2}$ are effective
$g$-factors for the corresponding spins (see for example \cite{DesSe06}), one
can show that the evolution operator $\hat{U}\left(  t\right)  $\ for the
equation (\ref{1}) can be reduced to an evolution operator $\hat{u}\left(
t\right)  $ (\ref{4a}) for the Schrödinger equation of a two-level system
\cite{BagBaGL07}. Such a reduction is given by the equation%
\begin{align}
&  \hat{U}\left(  t\right)  =\exp\left(  -\frac{i}{2}\left[  \left(
\Sigma_{3}+\rho_{3}\right)  \Gamma\left(  t\right)  +\Sigma_{3}\rho_{3}%
\Phi\left(  t\right)  \right]  \right)  M\left(  t\right)  ~,\nonumber\\
&  \Gamma\left(  t\right)  =\int_{0}^{t}B_{+}\left(  \tau\right)
\,d\tau\,,\;\Phi\left(  t\right)  =\int_{0}^{t}J\left(  t\right)
\,d\tau\,,\;B_{+}=\mu_{B}\left(  g_{1}B_{1}+g_{2}B_{2}\right)  \,,\nonumber\\
&  M=\left(
\begin{array}
[c]{cccc}%
1 & 0 & 0 & 0\\
0 & u_{11} & u_{12} & 0\\
0 & u_{21} & u_{22} & 0\\
0 & 0 & 0 & 1
\end{array}
\right)  ~, \label{2}%
\end{align}
where $\hat{u}\left(  t\right)  =||u_{ij}||$ obeys the Schrödinger equation
for the following two-level system (see \cite{BagGiBL05})%
\begin{align}
&  i\frac{d\hat{u}}{dt}=\left(  \boldsymbol{\sigma}\mathbf{\cdot K}\right)
\hat{u}\,,\;\hat{u}\left(  0\right)  =I\,,\nonumber\\
&  \mathbf{K}\left(  t\right)  =\left(  J\left(  t\right)  ,0,B_{-}\left(
t\right)  \right)  \,,\;B_{-}=\mu_{B}\left(  g_{1}B_{1}-g_{2}B_{2}\right)  \,.
\label{3}%
\end{align}
Thus, in the case under consideration, the four-level system problem\ is
reduced to solve the two-level system problem (\ref{3}) with an effective
magnetic field $\mathbf{K}$.

We can now use the expression (\ref{6a}) to construct exact solution for the
four-level system (\ref{1}). For an external field $\mathbf{K}$ in the form of
(\ref{3}) we have from (\ref{8b}),
\[
\frac{\dot{\alpha}}{2}\sin\theta\sin\varphi-\frac{\dot{\theta}}{2}\cos
\varphi=0~,
\]
and a solution in the form (\ref{6a}) can be construct for an external field
in the form%
\begin{equation}
K_{1}=J=\frac{\dot{\theta}}{2\sin\varphi}~,\ K_{3}=B_{-}=-\frac{\dot{\eta}%
}{2\tan\varphi}~,\ \eta=\ln\left(  \cos\varphi\sin\theta\right)  ~, \label{4}%
\end{equation}
with $\theta$ and $\alpha$\ any continuous time-dependent functions with
continuous derivative. So we can construct restricted in time interactions
$J$\ or fields difference $B_{-}$\ by choosing functions $\theta$ and $\eta$
that assume a constant value outside of a desired interval. Also in this case,
explicit examples can be constructed with the functions given in the preceding section.

\subsection{ An adiabatic variation of the field difference in each spin}

Although the expression (\ref{4}) allows to construct a variety of external
fields with some particular characteristic, the solution for a general field
can hardly be constructed in this manner. In this section we will analyze a
special case for a specific external parallel field (\ref{5})\ restricted in
time. Consider a four-level system in which the field difference $B_{-}$
(\ref{3}) varies adiabatically with time, i.e., a variation that met the
adiabaticity criteion \cite{BurLoDS99}, while the interaction function is
constant. Namely, we chose%
\begin{equation}
J=a\,,\;B_{-}\left(  t\right)  =c/\cosh\omega t\,, \label{2-1}%
\end{equation}
where $a,~c$ and $\omega$ are real constants. In practical application the
pulse applied to the system (e.g., two coupled quantum dots) needs to be
shorter than the decoherence time of the system. But such fast pulse can cause
a transition of the system to higher energy levels and, consequently, its
dynamic can no longer be described by the Hamiltonian (\ref{1b}). The
$c/\cosh\omega t$ dependence is the most adequate kind of a variation to avoid
this higher energy level transition \cite{BurLoDS99}. In addition, it is
reasonable to assume that if the only quantity that varies is $B_{-}$, and
$B_{+}\gg B_{-}$, the interaction function will remain constant \cite{BalGi07}%
. With regard to the variation of $B_{-}$, there are some proposals for the
application of localized magnetic fields \cite{GolLo02} and some techniques
that permit the manipulation of the $g$-factor by changing the size of the
dots or by the application of external electromagnetic fields
\cite{DesSe06,MedRiW03}.

From the previous Sect., we know that the evolution operator (\ref{2}) of a
four-level system with the parameters (\ref{2-1}) is expressed via an
evolution operator of a two-level system with effective field%
\begin{equation}
\mathbf{K}\left(  t\right)  =\left(  a,0,c/\cosh\omega t\right)  \,.
\label{2-2}%
\end{equation}
The exact solution for the evolution operator with such a field can be
constructed using our previous results \cite{BagGiBL05}. It has the form%
\begin{equation}
\hat{u}\left(  t\right)  =\frac{1}{\left\vert G_{1}^{0}\right\vert
^{2}+\left\vert G_{2}^{0}\right\vert ^{2}}\left(
\begin{array}
[c]{cc}%
G_{1}\left(  z\right)  & -\bar{G}_{2}\left(  z\right) \\
G_{2}\left(  z\right)  & \bar{G}_{1}\left(  z\right)
\end{array}
\right)  \left(
\begin{array}
[c]{cc}%
\bar{G}_{1}^{0} & \bar{G}_{2}^{0}\\
-G_{2}^{0} & G_{1}^{0}%
\end{array}
\right)  \,, \label{2-3}%
\end{equation}
where%
\begin{align}
&  G_{1}\left(  z\right)  =i\left(  2c+\omega\right)  z^{\mu}\left(
1-z\right)  ^{\nu}F\left(  \alpha,\nu;\gamma;z\right)  ~,\nonumber\\
&  G_{2}\left(  z\right)  =2az^{\mu+1/2}\left(  1-z\right)  ^{\nu}F\left(
\alpha,\nu+1;\gamma+1;z\right)  ~,\ G_{1,2}^{0}=G_{1,2}\left(  -1\right)
\,,\nonumber\\
&  z=\left(  \frac{e^{\varphi}+i}{e^{\varphi}-i}\right)  ^{2}\,,\;\varphi
=\omega t,\ \;\alpha=\gamma+\nu~,\nonumber\\
&  \mu=\frac{c}{2\omega},\;\nu=i\frac{\left\vert a\right\vert }{\omega
}\,,\;\gamma=\frac{1}{2}+2\mu~, \label{2-4}%
\end{align}
$F\left(  \alpha,\beta,\gamma,z\right)  $ is the Gauss hypergeometric
function, and complex conjugate quantities are designated by a bar above.

Substituting (\ref{2-4}) into (\ref{3}), we obtain%
\[
\hat{R}\left(  t\right)  =\exp\left(  \frac{iat}{2}\right)  \left(
\begin{array}
[c]{ccc}%
\exp\left[  -i\left(  at+\Gamma\left(  t\right)  \right)  \right]  &
\begin{array}
[c]{cc}%
0 & 0
\end{array}
& 0\\%
\begin{array}
[c]{c}%
0\\
0
\end{array}
& \hat{u}\left(  t\right)  &
\begin{array}
[c]{c}%
0\\
0
\end{array}
\\
0 &
\begin{array}
[c]{cc}%
0 & 0
\end{array}
& \exp\left[  -i\left(  at-\Gamma\left(  t\right)  \right)  \right]
\end{array}
\right)  ~,
\]
with $\hat{u}\left(  t\right)  $ given in (\ref{2-3}).

Thus, any transition amplitude for the four-level system can be calculated
with the help of the evolution operator. Let us, for example, calculate the
transition amplitude between the states $\left\vert \uparrow\downarrow
\right\rangle $ and $\left\vert \downarrow\uparrow\right\rangle $, which have
the form
\begin{equation}
\left\vert \uparrow\downarrow\right\rangle =\left\vert \uparrow\right\rangle
\otimes\left\vert \downarrow\right\rangle =\left(
\begin{array}
[c]{c}%
0\\
1\\
0\\
0
\end{array}
\right)  ,\;\;\left\vert \downarrow\uparrow\right\rangle =\left\vert
\downarrow\right\rangle \otimes\left\vert \uparrow\right\rangle =\left(
\begin{array}
[c]{c}%
0\\
0\\
1\\
0
\end{array}
\right)  ~. \label{2-6}%
\end{equation}
The transition between these states represents, in quantum computation, the so
called Swap operation and can be experimentally measured \cite{PetJoTLYLMHG05}%
. From the general expression (\ref{2}), we see that%
\begin{equation}
\left\langle \uparrow\downarrow\right\vert \hat{R}\left\vert \downarrow
\uparrow\right\rangle =\left\langle \uparrow\right\vert \hat{u}\left\vert
\downarrow\right\rangle ~,\ \left\langle \downarrow\uparrow\right\vert \hat
{R}\left\vert \uparrow\downarrow\right\rangle =\left\langle \downarrow
\right\vert \hat{u}\left\vert \uparrow\right\rangle ~,\ \ \left\vert
\uparrow\right\rangle =\left(
\begin{array}
[c]{c}%
1\\
0
\end{array}
\right)  ,\ \left\vert \downarrow\right\rangle =\left(
\begin{array}
[c]{c}%
0\\
1
\end{array}
\right)  ~. \label{l7}%
\end{equation}
Therefore, in the case of the Swap operation between the states (\ref{2-6}),
we need only to calculate matrix elements of the two-level system evolution
operator. One has also to stress that in this case the Swap operation does not
depend on the fields' sum $B_{+}$ .

Using (\ref{2-3}), we calculate the probability amplitude for the Swap
operation with the adiabatic variation (\ref{2-1}),%
\[
\left\vert \left\langle \downarrow\right\vert \hat{u}\left\vert \uparrow
\right\rangle \right\vert ^{2}=\frac{\left\vert G_{2}\left(  z\right)  \bar
{G}_{1}^{0}-\bar{G}_{1}\left(  z\right)  G_{2}^{0}\right\vert ^{2}}{\left(
\left\vert G_{2}^{0}\right\vert ^{2}+\left\vert G_{1}^{0}\right\vert
^{2}\right)  ^{2}}~.
\]

In order to use the adiabatic pulse to implement some quantum operations (like
the Swap or the XOR gate) the duration of the pulse needs to be shorter than
the dephasing time of the system. For example, in \textrm{GaAs} quantum dots
this time is about $10%
\operatorname{ns}%
$ \cite{PetJoTLYLMHG05}, which correspond to $\omega\simeq1%
\operatorname{GHz}%
$. In typical experimental conditions, we have fields of about $5%
\operatorname{T}%
$, $J=2\times10^{-3}%
\operatorname{eV}%
$ and, to satisfy the condition $B_{+}\gg B_{-}$, we can set the amplitude
$\left\vert B_{-}\right\vert =11%
\operatorname{mT}%
$. So, some characteristic values for our system are%
\[
\frac{\left\vert a\right\vert }{\omega}=\frac{\left\vert J\right\vert }%
{\hbar\omega}\simeq3~,\ \frac{c}{\omega}=\frac{\mu_{B}\left\vert
B_{-}\right\vert }{\hbar\omega}\simeq1~.
\]

In Figure 1 we have plots of the probability as a function of time for the
above values of the parameters. The first maximum occurs at $t=0.5%
\operatorname{ns}%
$ with a probability of $P=90\%$. For larger time, as the $\cosh^{-1}$
approaches to zero, this probability varies as $A_{1}\sin^{2}\left(
at\right)  +A_{2}$ where $A_{i}=A_{i}\left(  \omega,a,c\right)  $. The
amplitude $A_{1}$ decreases as $c$ increases while the shift $A_{2}$
increases. The functions $A_{i}$ change significantly with $\omega$ only for
$c>10a$.%

\begin{center}
\includegraphics[
height=1.4172in,
width=2.2059in
]%
{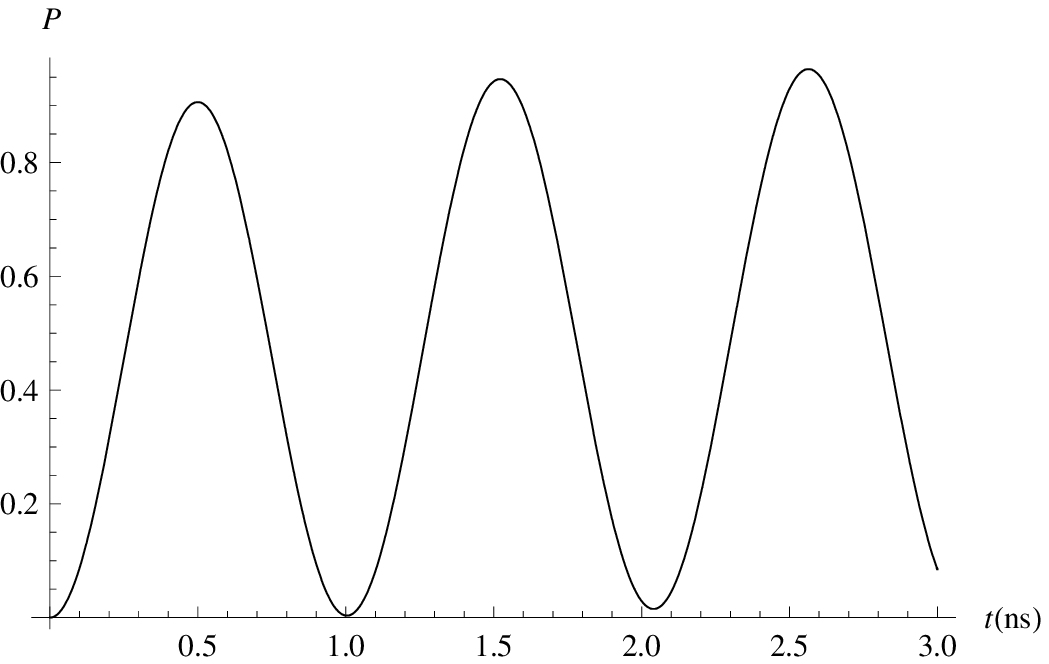}%
\\
{\small Figure 1 - Probability of the Swap operation as a function of time for
}$J=2\times10^{-3}~\operatorname{eV}${\small , }$\omega=1\operatorname{GHz}%
${\small  \ and }$B_{-}=11\times10^{-3}\operatorname{T}${\small .}%
\label{fig1}%
\end{center}
The dependence of the probability on the parameter $\omega$ become noticeable
for $c>4a$. In Figure 2 we plot this dependence for $a/c=2$ and $a/c=1/6$. The
parameter $\omega$ can be used to significantly attenuate the Swap transition
for values of $c>10a$.%
\begin{center}
\includegraphics[
trim=0.000000in 0.000000in -0.016786in 0.000000in,
height=1.3773in,
width=2.2449in
]%
{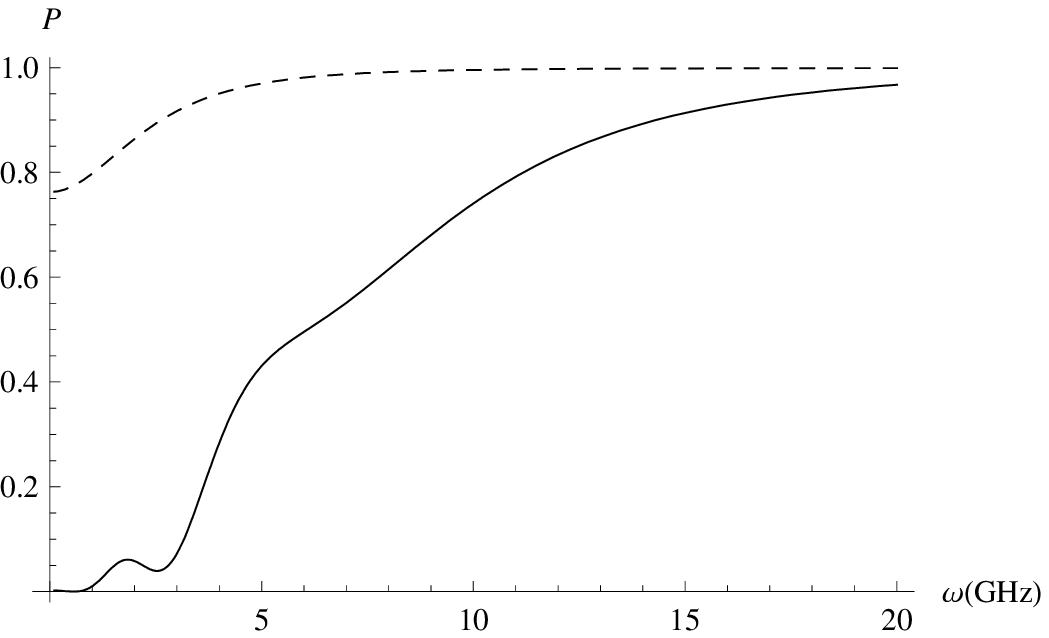}%
\\
{\small Figure 2 - Probability as a function of }$\omega${\small  \ for the
values }$c=1${\small  \ (dashed line) and }$c=12${\small  \ (solid line) in
}$t=0.8${\small  }$\operatorname{ns}${\small  \ and }$a=2${\small .}%
\label{fig2}%
\end{center}

A numerical study shows a strong dependence of the maximum values on the
parameter $a$. This fact can be used to measure the interaction $J$. In Figure
3 we plot the dependence of the probability on $J$. The attenuation of the
second maximum can be achieved by increasing the ratio $c/a$.%
\begin{center}
\includegraphics[
trim=0.000000in 0.000000in 0.004396in 0.000000in,
height=1.5326in,
width=2.4857in
]%
{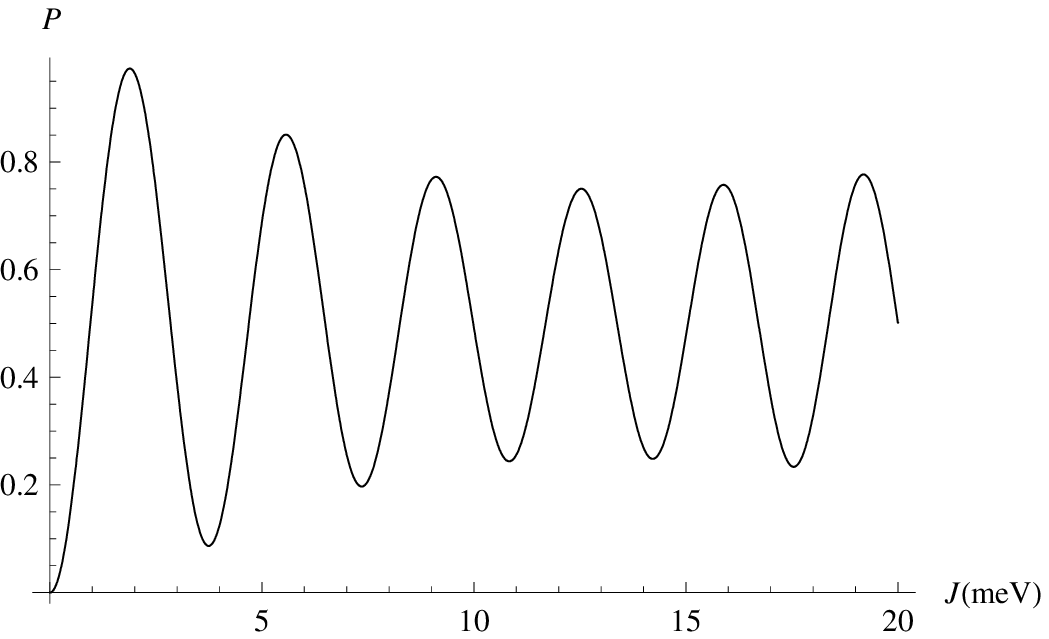}%
\\
{\small Figure 3 - Probability as a function of the interaction }$J${\small
\ for the }$c=30${\small , }$t=1${\small  }$\operatorname{ns}${\small  \ and
}$\omega=15\operatorname{GHz}${\small .}%
\label{fig3}%
\end{center}

\section{Final remarks}

We have described a method to construct some classes of exact solutions of the
one-spin equation and their respective external field configuration. Although
this method can not be used to solve the one-spin equation for a general
external field, it is very powerful in constructing solutions whose external
fields have some desired characteristic. As an application, we explicitly
construct the exact solutions for some restrict in time external fields. These
are a very important kind of fields, once the fields used in practical
application usually act in a finite time interval. After that, we show how
these results can be applied in the problem of two interacting spins subjected
to different magnetic fields, i.e., to solve the two-spin equation for
parallel fields. In this case, our method can be used to control, not only the
characteristics of the external magnetic fields, but also the behavior of the
interaction function between the spins. In a general manner, the exact
solutions of the two-spin equation have a wide application in the description
of two coupled quantum dots and, especially, in the construction of quantum
gates. In order to clarify this point, we show how an arbitrary exact
solution, not only the ones obtained with the method described here, can be
used to obtain the operational characteristics of a Swap operation. In this
analysis we chose the very important case of a restricted in time adiabatic
variation of the external magnetic field, and show explicitly the dependence
of the operational characteristics on the parameters of the external field.
Besides, we describe the behavior of the system using explicit values of the
parameters that can be obtained in experimental conditions. A graphical
analysis of this behavior is presented, from where we can see that the gate
can be better controlled if the experimental setup was assembled respecting
some specific relation between the parameters.

\begin{acknowledgement}
V.G.B. thanks grant SS-871.2008.2 of the President of Russia and RFBR grant
06-02-16719 for partial support; M.C.B. thanks FAPESP; D.M.G. thanks FAPESP
and CNPq for permanent support.
\end{acknowledgement}

\end{document}